\documentclass{statsoc}
\pdfoutput=1 

\usepackage[a4paper]{geometry}
\usepackage{graphicx}
\usepackage[textwidth=8em,textsize=small]{todonotes}
\usepackage{amsmath}
\usepackage{natbib}

\usepackage{longtable}
\usepackage{float}    
\usepackage{multirow}     
\usepackage{multicol}
\usepackage{longtable}
\usepackage{graphicx}     
\usepackage{mathtools}    
\usepackage{setspace}
\usepackage[labelfont=bf]{caption}
\usepackage{bm}   

\newcommand{\logit}{\mbox{logit}}
\newcommand{\argmax}{\operatornamewithlimits{argmax}}

\newcommand{\btheta}{{\bm \theta}}
\newcommand{\balpha}{{\bm \alpha}}
\newcommand{\bbeta}{{\bm \beta}}
\newcommand{\bb}[1]{\bm{#1}}

\usepackage{color}

\def\hei{\color{black}}
\newcommand{\yy}{\hei \rm}
\newcommand{\jj}{\hei \rm}

\title[AAA Design]{AAA: Triple-adaptive Bayesian designs for the identification of optimal dose combinations in dual-agent dose-finding trials}

\author{Jiaying Lyu}
\address{Department of Biostatistics, School of Public Health, Fudan University,
	Shanghai, 
	China.}
\author[Lyu {\it et al.}]{Yuan Ji}
\address{Program for Computational Biology \& Medicine, NorthShore University Health System,
	Evanston, 
	USA.}
\address{Department of Public Health Sciences, The University of Chicago, 
	Chicago, 
	USA. }
\email{koaeraser@gmail.com}
\author[Lyu {\it et al.}]{Naiqing Zhao}
\address{Department of Biostatistics, School of Public Health, Fudan University,
	Shanghai, 
	China.}
\author[Lyu {\it et al.}]{Daniel V.T. Catenacci}
\address{Department of Medicine, Section of Hematology \& Oncology,
	Chicago,
	USA.}
\address{University of Chicago Medical Center, 
	Chicago, 
	USA.}

\begin{document}
\begin{abstract}
	We propose a flexible design for the identification of optimal dose combinations in dual-agent dose-finding clinical trials. The design is called AAA, standing for three adaptations: adaptive model selection, adaptive dose insertion, and adaptive cohort division. The adaptations highlight the need and opportunity for innovation for dual-agent dose finding, and are supported by the numerical results presented in the proposed simulation studies. To our knowledge, this is the first design that allows for all three adaptations at the same time. We find that AAA improves the statistical inference, enhances the chance of finding the optimal dose combinations, and shortens the trial duration. A clinical trial is being planned to apply the AAA design. 
\end{abstract}
	
\maketitle

\section{Introduction}
Dual-agent dose-finding trials are becoming much more popular in oncology as more and new drugs become available.  
The traditional two-agent dose-finding trials often aim to capture the dose-toxicity relationship for the combinations and identify  one or more maximum tolerated dose combination (MTDC) of two agents. 
The MTDC is defined as the dose combination at which the probability that a patient experiences dose limiting toxicity (DLT) is less than a prespecified target rate $p_T$, which is usually \yy determined by physicians or clinical teams. \jj 
A large number of designs have been proposed to find the MTDC for trials with cytotoxic agents. \yy For example, \jj
\citet{Yuan2009} introduced Bayesian dose-finding approaches using copula regression models. 
\citet{braun2010hierarchical} developed a novel hierarchical Bayesian design accounting for patients heterogeneity, \yy and \jj  
\citet{hirakawa2013dose} developed a likelihood-based dose-finding method using a shrinkage logistic model. 
\citet{conaway2004designs} estimated the MTDC by determining the complete and partial orders of the toxicity probabilities \yy using \jj  nodal and non-nodal parameters. 
\yy Later, \citet{wages2011dose} applied model selection to estimate  possible complete orderings associated with the partial order based on the continual reassessment method (CRM). \jj 
As a review, \citet{hirakawa2015comparative} compared these five model-based dose-finding designs. 
They found that the performance of designs varied   depending  
on the dose matrix \yy and \jj the location and number of true MTDCs. 
More recently,  
\citet{mander2015product} published a curve-free method that relied on the product of independent beta probabilities. 
\citet{lin2016bayesian} developed a Bayesian optimal interval design for dual agents. 
\citet{sun2015two} proposed a two-stage adaptive algorithm based on modified biased coin design, 
and \citet{wages2016identifying} extended the CRM to identify an MTDC contour for dual agents. 

A key assumption to \yy all the works above \jj 
is the monotonicity of the dose-toxicity response 
and the dose-efficacy response, 
which is true 
in the case of cytotoxic agents \citep{LeTourneau2009}. 
As for many new cancer biological or immunological agents, such as the chimeric antigen receptors T-cell (CAR-T) therapies, such monotonic relationship may not be true, especially for dose-efficacy relationship \citep{Li2016}. For example, the dose-efficacy curves may follow a non-monotonic pattern, and efficacy may even decrease at higher dose levels \citep{Hoff2007}. 
Therefore, traditional dose-finding designs with a focus on finding the MTDC are not suitable for trials of non-cytotoxic 
agents. 
In contrast to the various literature for dual cytotoxic agents dose finding, there is a scarcity of designs for non-cytotoxic agents dose finding.
Instead of identifying the MTDC, one could consider the biologically optimal dose combination (BODC) for biological agents, the definition of which takes into account both efficacy and toxicity. 
\citet{wages2014phase} provided a phase I/II adaptive design to find a single dose combination with an acceptable level of toxicity that maximized efficacious response.   However, they assumed that the dose-toxicity and dose-efficacy relationships are monotonic among doses of one agent when the dose of other agent is fixed. 
\citet{Cai2014} proposed a novel dose-finding algorithm to encourage sufficient exploration of untried dose combinations in the two-dimensional space. 
\citet{guo2015bayesian} used isotonic regression to estimate partially stochastically ordered marginal posterior distributions of the efficacy and toxicity probabilities to estimate the BODC.

For dual-agent trials, due to the challenges in capturing the proper therapeutic range for the dose levels of both agents, the BODC might locate outside the candidate dose range or sandwiched by existing dose combinations. Therefore, a design capable of extrapolating or interpolating a new dose combination when candidate dose combinations are deemed suboptimal can drastically improve one's chance to identify better dose combinations. To this end, \citet{hu2013adaptive} considered an adaptive dose insertion scheme to allow new doses to be inserted during the course of a dose-finding trial. Later, \citet{chu2016adaptive} introduced an extended version. Both methods only consider toxicity outcomes. \citet{guo2015teams} proposed a toxicity- and efficacy- based dose insertion design with adaptive model selection (TEAMS) for single-agent trials, and illustrated the importance of correct model specification for dose insertion. They showed that in order to insert the right doses, the dose-efficacy relationship must be properly identified.  

In this paper,  we extend the idea of TEAMS to dual agents  and   propose the AAA \yy (triple A) \jj design. 
The AAA design is named after three adaptive features. First, to describe the appropriate dose-efficacy curve, we present an adaptive   Bayesian  model selection procedure based on median posterior probability models \citep{barbieri2004optimal} that allows the dose-efficacy model to vary between the monotone pattern and non-monotone pattern. Second, with the correct models being selected, we propose adaptive dose insertion allowing new dose combinations to be extrapolated or interpolated during the course of the trial. Last, 
importantly and innovatively, we consider adaptive cohort division (ACD) and allow multiple cohorts of patients to be   enrolled simultaneously during the course of the trial. We show that ACD accelerates trial conduct and shortens trial duration.  

We consider a conceived clinical study at The University of Chicago involving a MEK inhibitor and a PIK3CA inhibitor, both with four doses at their regular monotherapy dose, two lower doses and one higher dose. This Phase I dose finding study  will enroll late-stage caner patients with a primary endpoint aiming to improve efficacy rate from 5\% to 30\% with the optimal tolerated dose combination. We will use this study as the basis for our numerical studies later. 

The remainder of this article is organized as follows. In Sections \ref{Methods} and \ref{design}, we describe the probability model and the AAA design. In Section \ref{results}, using the phase I trial we examine the operating characteristics of the AAA design through simulation studies. To evaluate the time reduction by using \yy ACD,   we examine the duration of the trial in the simulated trials. \jj We conclude with a brief discussion in Section \ref{discussion}.

\section{Methods} \label{Methods}
\subsection{Dose-response models} \label{prob}
Consider a trial combining $J$ doses of agent A, denoted by $x_{a,1}<\cdots<x_{a,J}$, and 
$K$ doses of agent B, denoted by $x_{b,1}<\cdots<x_{b,K}$, for dose finding. Without loss of generality, we assume $J\geq K$ and that the dosage 
values of the $x_{a,j}$'s and $x_{b,k}$'s have been standardized to have mean 0 and standard deviation 0.5. Let $\bm{x}_{jk}=(x_{a,j},x_{b,k})$ denote the combination of dose   levels $j$ and $k$, 
and let $p(\bm{x}_{jk})$ and $q(\bm{x}_{jk})$ denote probabilities of the toxicity event and efficacy event for the dose combination $(x_{a,j},x_{b,k})$, respectively, for $j=1,2,\ldots,J$, and $k=1,2,\ldots,K$.

Assume that $p(\bm{x})$ follows a linear logistic model and $q(\bm{x})$ follows a quadratic logistic model in order to incorporate a non-monotone pattern in the dose-efficacy model
\begin{eqnarray}
\logit\{p(\bm{x})\} &=& \alpha_0+\alpha_1 x_{a}+\alpha_2 x_{b}, \label{eq:tox} \\
\logit\{q(\bm{x})\} &=& \beta_0+\beta_1 x_{a}+\beta_2 x_{b}+\beta_3 x_{a}^2+\beta_4 x_{b}^2, \label{eq:eff}
\end{eqnarray}
where $\alpha_1>0$, $\alpha_2>0$, and $\bm{x}=(x_a,x_b)$ is the vector of the dose combination. \yy Later we briefly discuss adding an interaction term $\beta_5 x_a x_b$ in the last section. \jj Denote $\balpha = (\alpha_0,\alpha_1,\alpha_2)^\prime$ and $\bbeta = (\beta_0,\beta_1,\beta_2,\beta_3,\beta_4)^\prime$, the vector of regression parameters in the dose-toxicity model (\ref{eq:tox}) and dose-efficacy model (\ref{eq:eff}), respectively. 
Here, we use a working model and assume that the binary outcomes of toxicity and efficacy are independent. This working independence between efficacy and toxicity outcome in dose-finding designs has been extensively discussed in the literature \citep{Cai2014,Ivaova2009}. We also assume that toxicity is monotone with the dose as a conservative choice of model. In other words, $\alpha_1 > 0 $ and $\alpha_2 > 0 $ in model (\ref{eq:tox}).  

\subsection{Utility function and definition of BODC} 
\begin{figure}
	\begin{center}
		\centerline{\includegraphics[width=1\textwidth]{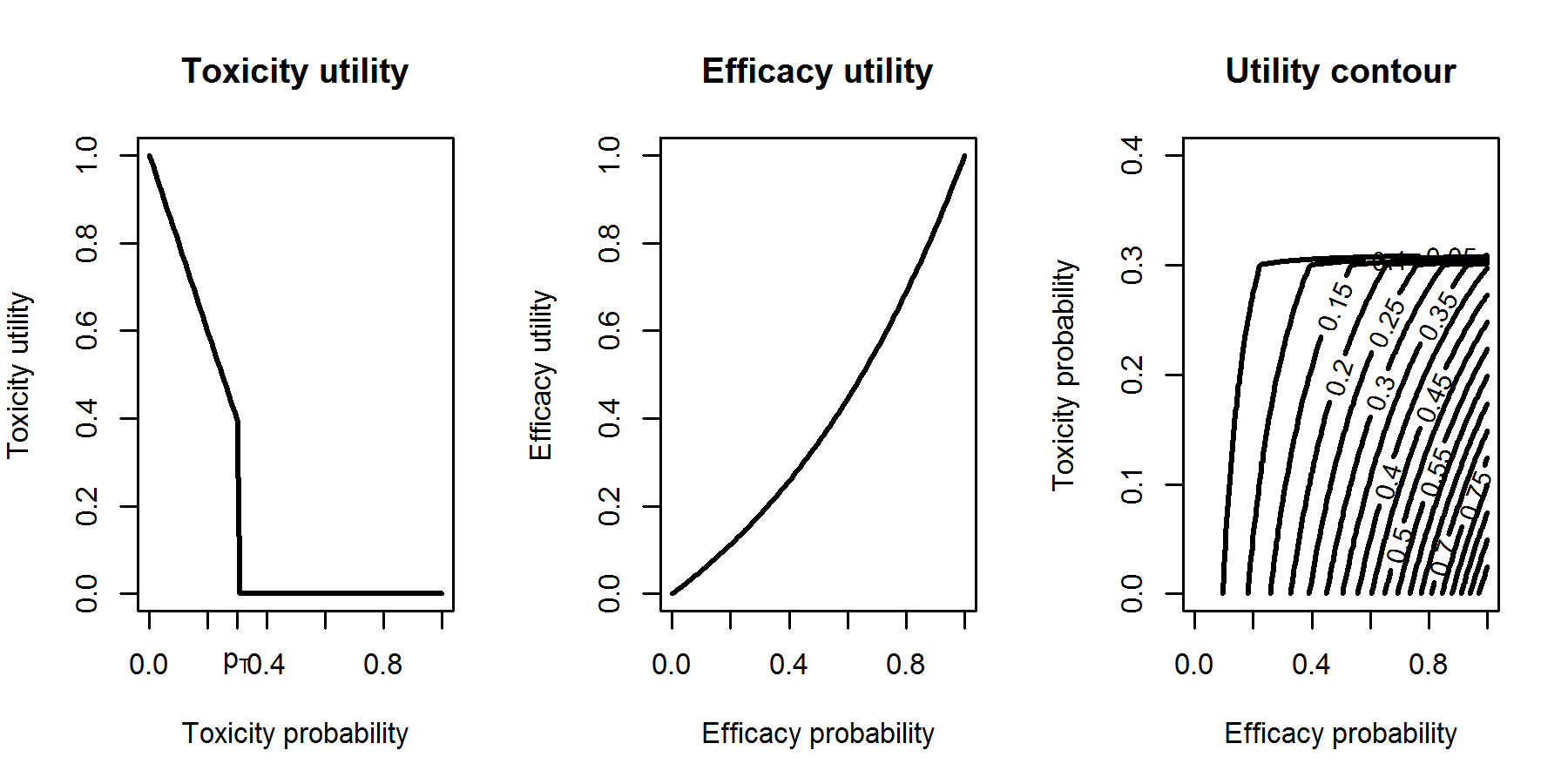}}
		\caption{Utility functions: From left to right are utility for safety (truncated at $p_T$ and sharply decreases to 0), utility for efficacy and the overall utility contours.
		} \label{fig:utility}
	\end{center}
\end{figure}

Utility-based decision criteria have been adopted \yy frequently \jj in recent dose-finding trials \citep{thall2012adaptive,lee2015bayesian,quintana2016bayesian,Li2016}. In this paper, we construct utility functions for 
dose safety and efficacy evaluation. 
Denote $U_T(p(\bm{x}),\eta_0)$ and $U_E(q(\bm{x}),\boldsymbol{\tilde{\eta}})$ the utility for safety and efficacy at dose combination $\bm{x}=(x_a,x_b)$, respectively, and define
\begin{eqnarray}
U_T(p(\bm{x}),\eta_0) &=&
\begin{cases}
1-\frac{1-\eta_0}{p_T}p(\bm{x}), & p(\bm{x}) \in (0,p_T], \\
0, & p(\bm{x}) \in (p_T,1),
\end{cases} \label{eq:utp}\\
U_E(q(\bm{x}),\boldsymbol{\tilde{\eta}}) &=& \eta_1 \cdot \exp \{\eta_2  \cdot q(\bm{x})\}+\eta_3, \mbox{~~~~~~} \eta_2>0, \label{eq:utq}
\end{eqnarray}
\yy where $\boldsymbol{\tilde{\eta}}=(\eta_1,\eta_2,\eta_3)^\prime$.   See Figure \ref{fig:utility} for an illustration.  \jj 
Here, the utility for safety 
$U_T$ in 
\eqref{eq:utp} is a truncated linear decreasing function with $p(\bm{x})$;
we assume that the toxicity utility $U_T$ decreases with toxicity probability and drops to 0 if $p(\bm{x})>p_T$, i.e., there is no utility when toxicity probability $p(\bm{x})$ is larger than $p_T$. 
Usually $p_T$ is around 0.3 for oncology trials. The utility for efficacy $U_E$ in \eqref{eq:utq} follows an exponential function with parameters $\eta_1$, $\eta_2$ and $\eta_3$, where $(\eta_1+\eta_3)$ decides the utility value when there is no efficacy and $(\eta_1,\eta_2)$ decide how fast utility increases when efficacy probability $q(\bb{x})$ increases. \yy Combining \jj  the utilities for toxicity and efficacy, we define the overall utility score as
\begin{equation}
U(\bm{x},\bb{\eta}) = U_T(p(\bm{x}),\eta_0) \cdot U_E(q(\bm{x}),\boldsymbol{\tilde{\eta}}), \label{eq:utility}
\end{equation}
where $\bb{\eta}=(\eta_0,\eta_1,\eta_2,\eta_3)^\prime$. 


Then the biologically optimal dose combination (BODC), $\bm{x}_{opt}=(x_{a,opt},x_{b,opt})$, is defined as the dose combination that maximizes the utility function, i.e. 
\begin{equation*}
\bm{x}_{opt}(\btheta) = \argmax_{\bm{x}}U(\bm{x},\btheta).
\end{equation*}


To specify the unknown values $(\eta_0,\eta_1,\eta_2,\eta_3)$, 
we follow a procedure suggested by \citet{thall2004dose}. We elicit with physicians two pairs of toxicity-efficacy trade-off, $(0,q_1^*)$ and $(p_T,q_2^*)$, that have the same utility value $U^*$, say $U^*=0.3$. For example, $q_1^*=0.1$, $q_2^*=0.3$. This gives the two equations: $U_T(0,\eta_0) \cdot U_E(q_1^*,\boldsymbol{\tilde{\eta}})=U^*$ and $U_T(p_T,\eta_0) \cdot U_E(q_2^*,\boldsymbol{\tilde{\eta}})=U^*$. In addition, $U_T$ and $U_E$ must have the same scale $(0,1)$, which implies that 
1) $U_E(q(\bm{x})=0,\boldsymbol{\tilde{\eta}})=\eta_1 + \eta_3 = 0$; and 2) $U_E(q(\bm{x})=1,\boldsymbol{\tilde{\eta}})=\eta_1 \cdot \exp (\eta_2) + \eta_3 = 1$. Therefore, we have a 
set of four nonlinear equations and four unknown parameters. The numerical solution of $\hat{\bb{\eta}}$ can be easily solved \yy numerically. \jj 

\subsection{Adaptive model selection for the dose-efficacy model} \label{AMS}
Efficacy could be either monotone or non-monotone with dose combination, depending on many factors such as the pharmacology and mechanism of action of the drug. Proposing adaptive model selection, we allow adaptation in the model choice throughout the trial. Briefly, when the number of explored dose combinations or the sample size is small, a simpler model, such as a linear logistic model, may fit the data better to avoid the wrong estimation of dose-response curve due to the model mis-specification. 
As the trial proceeds, more dose combinations are explored, 
more patient data are accumulated, and more complex models such as a non-monotone quadratic logit model might be beneficial to obtain better estimates \citep{guo2015teams}. 

Consider a 
model selection framework  
for the 
efficacy  
regression coefficients in \eqref{eq:eff} as follows: 
\begin{align*}
M_1: \beta_3 = \beta_4=0; & & M_2: \beta_3\neq 0,  \beta_4=0; & & M_3: \beta_3=0, \beta_4 \neq 0; & & M_4: \beta_3 \neq 0, \beta_4 \neq 0.
\end{align*}
Similar to \citet{guo2015teams}, we adopt the inverse moment priors (iMOM) \citep{Johnson2010} on $\beta_3$ under $M_2$ and $M_4$, $\beta_4$ under $M_3$ and $M_4$,  in which cases either or both of them are assumed to be non-zero. 
The iMOM prior has 
no probability mass at the null point $(\beta_i=0,i=3,4)$ and takes the form
\begin{align*}
\pi(\beta_i|M_l) = 
\frac{k\tau^{\nu/2}}{\Gamma(\nu/2k)}|\beta_i|^{-(\nu+1)}\exp\left(\frac{\tau^k}{\beta_i^{2k}}\right), \mbox{~~~~~~} (i,l) \in \{(3,2),(3,4),(4,3),(4,4)\}
\end{align*}
for $k,\nu,\tau>0$. The choice of iMOM prior is discussed in detail in Appendix A. The prior  when $\beta_3$ \yy and \jj $\beta_4$ equals zero  is simply a point mass at zero, i.e.
$$\pi(\beta_i|M_l)=\bm{1}\{\beta_i=0\}, \mbox{~~~~~~} (i,l) \in \{(3,1),(3,3),(4,1),(4,2)\},$$
where $\bm{1}\{\cdot\}$ is the indicator function. With no evidence favoring any of the hypotheses over the others a priori, we take $P(M_1)=P(M_2)=P(M_3)=P(M_4)=1/4$. 

In the model selection, we compute $P(M_l|Data)$, the posterior probability of each model and  
select the \textit{median probability model (MPM)} to be the 
dose response model. MPM 
is defined as the model consisting of those variables which have overall posterior probability greater than or equal to $1/2$ \citep{barbieri2004optimal}. Specifically, in this paper, denote $p_3$ and $p_4$ the posterior inclusion probability for the quadratic terms $x_{a}^2$ and $x_{b}^2$, respectively, and define
\begin{align}
p_{3}=P(M_2|Data)+P(M_4|Data), \label{eq:q3} \\
p_{4}=P(M_3|Data)+P(M_4|Data), \label{eq:q4}
\end{align}
which are also the overall posterior probability that $\beta_3 \neq 0$ and $\beta_4 \neq 0$, respectively. The posterior probability of model $M_l$, $P(M_l|Data)$ in \eqref{eq:q3} and \eqref{eq:q4}, $l=2,3,4$, is given by
\begin{eqnarray*}
	P(M_l|Data) = \frac{P(Data|M_l)P(M_l)}{\sum_{l=1}^{4}P(Data|M_l)P(M_l)},
\end{eqnarray*}
where $P(Data|M_l)$ is the marginal distribution of the data under the prior of model $M_l$, given by
\begin{eqnarray*}
	P(Data|M_l) = \int \pi(\balpha,\bbeta_l|M_l) \mathcal{L}(Data|\balpha,\bbeta_l,M_l) d\balpha d\bbeta_l. 
\end{eqnarray*}
Here $\mathcal{L}(Data|\balpha,\bbeta_l,M_l)$ is the likelihood function under model $M_l$, $l=1,2,3,4$, $\bbeta_1=\{\beta_0,\beta_1,\beta_2,\beta_3=\beta_4=0\}$, $\bbeta_2=\{\beta_0,\beta_1,\beta_2,\beta_3,\beta_4=0\}$, $\bbeta_3=\{\beta_0,\beta_1,\beta_2,\beta_3=0, \beta_4\}$ and $\bbeta_4=\{\beta_0,\beta_1,\beta_2,\beta_3,\beta_4\}$. Since the integral does not have a closed form, numerical integration such as Monte Carlo integration is applied. Specifically, we use the harmonic mean of likelihood values here \citep{Kass1995}. Let $\bbeta_l^{(1)}, \bbeta_l^{(2)}, \dots, \bbeta_l^{(B)}$ be a Markov chain Monte Carlo (MCMC) sample from the posterior distribution of $\bbeta_l$ under model $M_l$; suppressing terms related to $\boldsymbol{\alpha}$ it can be shown that
\begin{eqnarray*}
	\int
	\pi(\boldsymbol{\beta_l}|M_l)\mathcal{L}(Data|\boldsymbol{\beta}_l,M_l)d\boldsymbol{\beta}_l\approx
	\left\{\frac{1}{B}\sum_{b=1}^B\mathcal{L}(Data|\boldsymbol{\beta}_l^{(b)},M_l)^{-1}\right\}^{-1},
\end{eqnarray*}
for $l =1,2,3,4$. \citet{Kass1995} showed that the harmonic mean approach is more efficient than directly sampling from the prior, especially when the likelihood function is highly concentrated in an area with low prior probabilities.

We perform model selection based on posterior inclusion probability, $p_3$ and $p_4$ (Table \ref{tab:mpm}). For instance, If  $p_3 \geq 1/2$ and $p_4 \geq 1/2$, the quadratic terms of both agents, $x_{a}^2$ and $x_{b}^2$, are included in the model \eqref{eq:eff}, i.e. $\beta_3, \beta_4 \neq 0$.
Therefore, we select the model $M_4$. 


\subsection{\yy Adaptive \jj dose combination insertion}
 The therapeutic window of two different drugs is often complex and difficult to delineate. In a trial that prespecifies a set of dose combinations for investigation,  a new dose combination should be inserted when the BODC, $\bb{x}_{opt}$, is distant to all the existing dose combinations in the trial.  This is our second proposed adaptation.  Mathematically, we propose an activation rule for triggering the dose-insertion procedure. 
Let $\mathcal{R}_C(\bb{x}_{opt})$ represent the $C\%$ (e.g., $C=90$) 
posterior 
credible circular region of $\bb{x}_{opt}$, 
defined as follows:
\begin{equation*}
\mathcal{R}_C(\bb{x}_{opt},r) = \{
(x_{a,opt},x_{b,opt}):  Pr \{(x_{a,opt}-x_{a,0})^2+(x_{b,opt}-x_{b,0})^2 \leq r \mid Data \}=C \% \}, 
\end{equation*} 
where $(x_{a,0},x_{b,0})$ and $r$ are the center and the radius of the circular region, respectively. Define $A$ as the indicator of dose insertion,
\begin{align}
A =
\begin{cases}
1,  & \mbox{~~~if~~~}\mathcal{R}_C(\bb{x}_{opt},r) \bigcap \{(x_{a,j},x_{b,k}): j=1,\ldots,J, k=1,\ldots,K\} = \emptyset,   \\
0,  & \mbox{~~~if~~~}\mathcal{R}_C(\bb{x}_{opt},r) \bigcap \{(x_{a,j},x_{b,k}): j=1,\ldots,J, k=1,\ldots,K\} \neq \emptyset, 
\end{cases}
\label{doseinsertion}
\end{align}
where $\emptyset$ denotes the empty set. When $A=1$, the credible region does not cover any existing dose combinations, and the dose-insertion procedure is activated. Otherwise, the trial proceeds by treating the next cohort at one of the existing dose combinations. 

\subsection{Adaptive cohort division}
 Adaptive cohort division (ACD) is the third and an innovative adaptation.   When two or more doses are considered similarly desirable for the next cohort of patients based on the collected data,  the proposed AAA design allows patients to be enrolled simultaneously in parallel cohorts. 

The main idea is as follows. When we encounter a toxic dose combination during the trial,  \yy a de-escalation is needed that decreases the dose level of either drug. To increase efficiency, \jj we propose to de-escalate to two untried lower dose combinations with parallel patient enrollment at both dose combinations.   That is, we open two cohorts concurrently in this case. 
Cohorts are collapsed if a new dose combination is inserted, in which case the new single cohort will be enrolled at the inserted dose, or the multiple cohorts all point to the same dose combination for future patients. 

  Because of the ACD procedure, multiple cohorts can be enrolled at the same time. Some cohorts might finish enrollment and follow up faster than others.  \yy When a cohort finishes follow up, the efficacy and toxicity  response data of the patients in the cohort are observed.  At this point, \jj a decision must be made as to the next dose combination for future patients. However, at that moment  there might be other cohorts still enrolling, in which case \yy some patients might still be followed without response data while others might have completed  follow up with outcomes. \jj  To fully use   the existing information, we include the patients with complete data in all cohorts in the inference and decision making.  In other words, \yy we make a decision on the next dose combination based on the response data  from all completers from all cohorts. \jj This achieves faster enrollment and exploration of the new dose combinations, thereby shortening trial duration.   


\subsection{Likelihood and prior specification} \label{likelihood}
Let $y_{jk}$, $z_{jk}$ and $n_{jk}$ be the numbers of toxicity responses, efficacy responses and total patients treated at dose combination $(x_{a,j},x_{b,k})$  when a cohort completes follow up during the trial, for $j=1,2,\cdots, J$, and $k=1,2,\cdots,K$. Note these numbers include all completers in all cohorts.  For the observed $Data \equiv \{(y_{jk},z_{jk},n_{jk}), j=1,2,\cdots, J, k=1,2,\cdots,K\}$, the likelihood function under model $M_l$ is the product of the binomial densities, i.e. 
\begin{equation*}
\mathcal{L}(Data|\balpha,\bbeta_l,M_l) \propto 
\prod_{j=1}^{J} \prod_{k=1}^{K} p(\bm{x}_{jk}|\balpha)^{y_{jk}} \{1-p(\bm{x}_{jk}|\balpha)\}^{n_{jk}-y_{jk}} 
q(\bm{x}_{jk}|\bbeta_l)^{z_{jk}} \{1-q(\bm{x}_{jk}|\bbeta_l)\}^{n_{jk}-z_{jk}}
\end{equation*}
where $l=1,2,3,4$ index four different models. Denote $\pi_E(\bbeta_l|M_l)$ and $\pi_T(\balpha)$ the priors for $\bbeta_l$ and $\balpha$. Assuming the prior independence between $\bbeta_l$ and $\balpha$, the joint conditional posterior under $M_L$ is given by 
\begin{equation*}
\pi(\btheta_l|Data,M_l) \propto \mathcal{L}(Data|\balpha,\bbeta_l,M_l) \pi_E(\bbeta_l|M_l)  \pi_T(\balpha),
\end{equation*}
where $\btheta_l=(\balpha^\prime,\bbeta_l^\prime)^\prime$. 

For the prior specification of  parameters in the efficacy model \eqref{eq:eff} other than $\beta_3$ and $\beta_4$,  we use a weakly informative default prior for $\beta_0$, $\beta_1$ and $\beta_2$, recommended by \citet{gelman2008weakly}. That is, $\beta_0 \sim Cauchy(0,10)$, and $\beta_1,\beta_2 \sim Cauchy(0,2.5)$, where $Cauchy(c,d)$ denotes a Cauchy distribution with the center parameter $c$ and the scale parameter $d$. 
These weakly informative and appropriately regularized priors improves the estimation stability and still ensures that the data are able to dominate the priors \citep{gelman2008weakly}. For the iMOM priors for $\beta_3$ and $\beta_4$, we use the default values for $k$ and $\nu$: $k=\nu=1$, recommended by \citet{Johnson2010}.  With respect to the choice of parameter $\tau$, we provide a function of the dose level $\tau=f(x_1,x_2,\cdots,x_J)$ to set $\tau$ for different scenarios.
See detail in Web Appendix A. 

For the toxicity model \eqref{eq:tox}, we also adopt the weakly informative default prior $Cauchy(0,10)$ for intercept $\alpha_0$. We assign $\alpha_1$ and $\alpha_2$ independent gamma distributions with the shape parameter of $0.5$ and the rate parameter of $0.5$. 
This gives mean 1 and variance 2.

\section{Trial design}  \label{design}
\subsection{Overview}
The proposed dose-finding design consists of two stages. Stage I is a run-in period, in which we escalate the dose along the diagonal of the dose combination matrix in order to explore the dose-combination space quickly and collect preliminary data for stage II. This is similar to \citet{Cai2014}. 
Stage I stops if 
we reach the highest dose combination $(x_{a,J},x_{b,K})$ or  if  we encounter a dose combination that violates the safety requirement, i.e., $Pr \{p(\bb{x}_{jk})>p_T | Data\}>\xi$, where $\xi$ is a prespecified safety cutoff probability and often set to be close to 1, e.g. $\xi=0.95$. 
If the dose matrix is not square (i.e., $J>K$), after first escalating the dose along the diagonal to the $(x_{a,K},x_{b,K})$, we escalate the dose by holding the dose level of agent $B$ at $K$ and increasing the dose level of agent $A$ from $(x_{a,K},x_{b,K})$ to $(x_{a,K+1},x_{b,K})$ and so on. 
After stage I, the trials enters stage II, adaptive dose finding.

In stage II, we 
apply  
\yy the toxicity and efficacy probability models for inference, the utility function for dose assessment, \jj
 and the three adaptive procedures (model selection, dose insertion and cohort division)  
in Section \ref{Methods} for 
adaptive 
dose finding. 
A simple flow chart in Figure \ref{fig:flowchart} depicts the flow of stage II in the AAA design. 
Specifically, once a cohort of patients completes follow up in the trial,  
we update the recorded outcome data from existing doses and enrolled patients, generate MCMC posterior samples of the parameters under $M_1$, $M_2$, $M_3$ and $M_4$ respectively, denoted by $\{\btheta_l^{(b)}, b=1,2,\cdots,B\}$, $l=1,2,3,4$, and carry out the adaptive model selection based on the MPM using the updated data. Suppose $M_{l^*}$ is selected, we obtain an MCMC posterior sample of $\btheta_{l^*}$ under the selected $M_{l^*}$. For each simulated values $\btheta_{l^*}^{(b)}$ from the $b$-th MCMC iteration, $b=1,2,\cdots,R$, we maximize the utility function $U(\bb{x},\btheta_{l^*}^{(b)})$ with respect to dose combination $\bb{x}$, to obtain a posterior sample of BODC, i.e. 
\begin{equation}
\hat{\bb{x}}_{opt,l^*}^{(b)}=(\hat{x}_{a,opt,l^*}^{(b)},\hat{x}_{b,opt,l^*}^{(b)})=\argmax_{\bb{x}}
U(\bb{x},\boldsymbol{\theta}_{l^*}^{(b)}), \quad l^* \in \{1, 2, 3, 4\},
\label{eq:bodc}
\end{equation}
for which the posterior mean  of BODC is estimated to be 
\begin{equation}
\hat{\bb{x}}_{opt} = (\hat{x}_{a,opt},\hat{x}_{b,opt}) = \frac{\sum_{b=1}^B \hat{\bb{x}}_{opt,l^*}^{(b)}}{B}.
\label{meanbodc}
\end{equation}

\subsection{Deciding the next dose combination}
We  first  compute the dose-insertion activator $A$ using the posterior sample of BODC $\{\hat{\bb{x}}_{opt,l^*}^{(b)}: b=1,2,\cdots,B\}$, and let 
\begin{eqnarray}
\hat{A} &=& \bb{1} \left\{ Pr \left\{ (x_{a,opt,l^*}-\hat{x}_{a,opt})^2+(x_{b,opt,l^*}-\hat{x}_{b,opt})^2 \leq \hat{r} \right\} > C \% \right\} \nonumber \\
&\approx & \bb{1} \left\{ \frac{1}{B} \sum_{b} \bb{1} \left\{  x_{a,opt,l^*}^{(b)}-\hat{x}_{a,opt})^2+(x_{b,opt,l^*}^{(b)}-\hat{x}_{b,opt})^2 \leq \hat{r}  \right\}
> C \% \right\}, \label{doseinsertion2} 
\end{eqnarray} 
where $\bb{1}\{\cdot\}$ is the indicator function and $\hat{r}$ is the minimum Euclidean distance among the distances between the center and the existing dose combinations, denoted by 
$$\hat{r} = \min_{j,k} \left \{\sqrt{(x_{a,j}-\hat{x}_{a,opt})^2+(x_{b,k}-\hat{x}_{b,opt})^2} \right \}.$$ 
We can easily see that   $\hat{A}$ in \eqref{doseinsertion2} is a posterior estimate of \eqref{doseinsertion}. 

\begin{figure}
	\begin{center}
		\centerline{\includegraphics[width=0.6\textwidth]{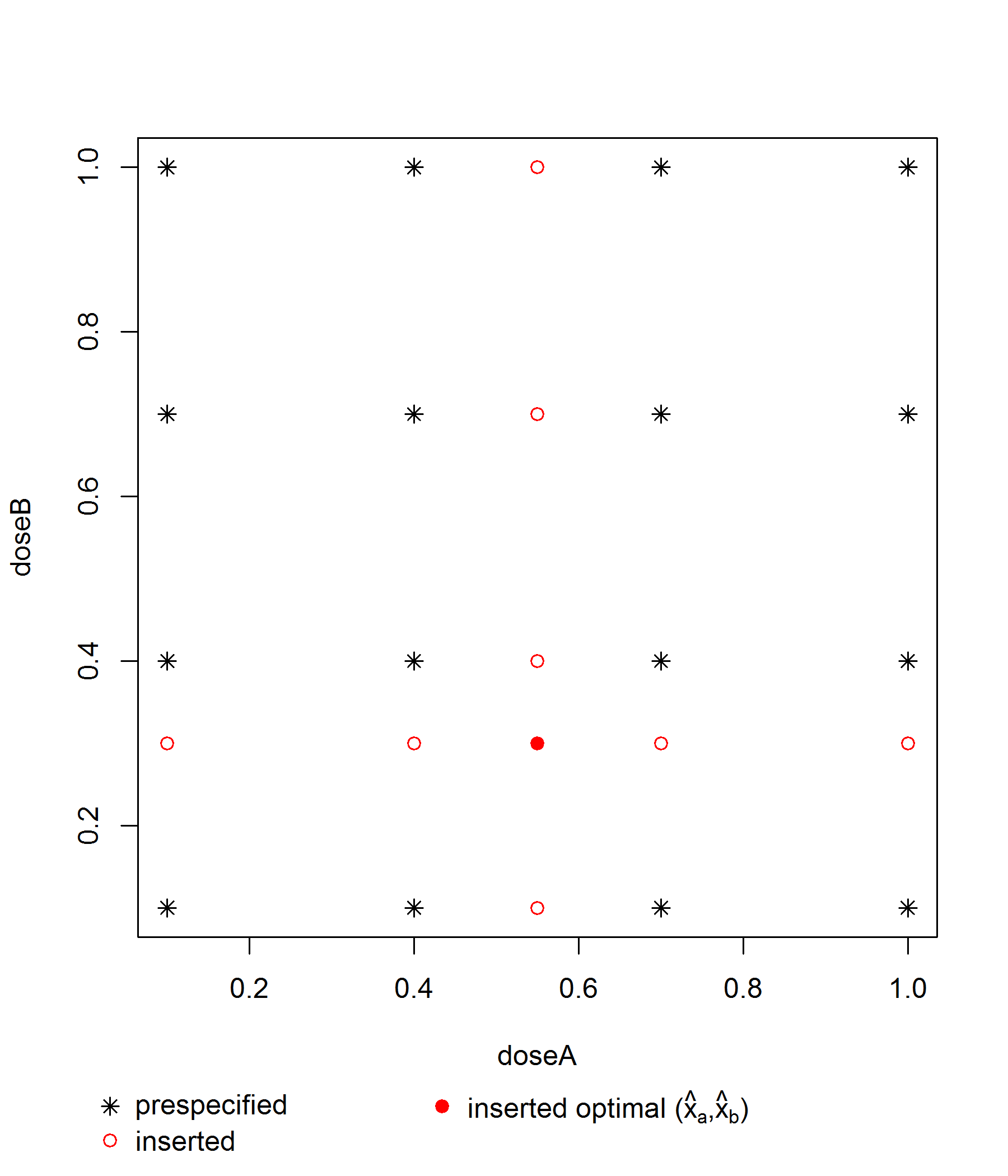}}
		\caption{\small The procedure of dose insertion. Black dots: prespecified dose combinations, red dots: inserted dose combinations.
		} \label{fig:insert}
	\end{center}
\end{figure}

\begin{figure}
	\begin{center}
		\centerline{\includegraphics[width=1\textwidth]{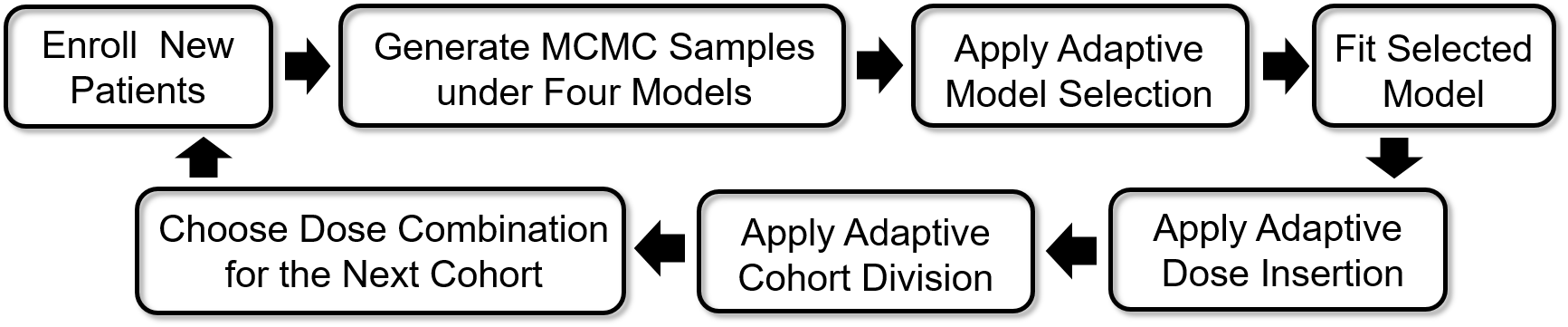}}
		\caption{\small A simple flow chart for stage II in the AAA design.
		} \label{fig:flowchart}
	\end{center}
\end{figure}

If dose insertion is needed, i.e., $\hat{A}=1$, we insert the new dose combination $(\hat{x}_{a,opt},\hat{x}_{b,opt})$ and two 
sets of new dose combinations  
$$\{(x_{a,1},\hat{x}_{b,opt}), \cdots, (x_{a,J},\hat{x}_{b,opt})\} \quad \mbox{and} \quad  \{(\hat{x}_{a,opt},x_{b,1}), \cdots, (\hat{x}_{a,opt},x_{b,K})\}$$ into the dose combination matrix, as shown in Figure \ref{fig:insert}, and assign the next cohort to the new dose combination $(\hat{x}_{a,opt},\hat{x}_{b,opt})$. 

If dose insertion is not needed, i.e., $\hat{A}=0$, we assign the next cohort of patients according to the   utility of the existing dose combinations.  
Let $N$ denote the prespecified maximum sample size, $N_1$ denote the number of patients in stage I, and $N_2=N-N_1$ be the total number of patients available for stage II. Given the current dose combination $\bb{x}_{jk}=(x_{a,j},x_{b,k})$, we define 1-degree admissible dose set, denoted by $\mathcal{A}_1$, as dose combination $\bb{x}_{j^\prime k^\prime}$, whose dose levels are different from $\bb{x}_{jk}$ no more than 1 level and satisfy the safety requirement, i.e., $\mathcal{A}_1=\{ \bb{x}_{j^\prime k^\prime}: \left| j^\prime -j\right|\leq 1, \left| k^\prime -k\right|\leq 1, Pr\{ p(\bb{x}_{j^\prime k^\prime})>p_T | Data\} \leq \xi \}$, where $\xi$ is close to 1. Then  the next dose combination is decided based on the following algorithm:  

\bigskip

\noindent {\bf If $(x_{a,j},x_{b,k})$ is considered safe, } i.e., $Pr\{ p(\bb{x}_{jk})>p_T | Data\} \leq \xi$, where $\xi$ is close to 1, assign patients as follows: 

		[a.] Based on the accumulated trial data, determine dose set $\mathcal{A}_1$. 

		[b.] Among the dose  combinations   in $\mathcal{A}_1$, compute the posterior mean utility for each combination, i.e., $\bar{U}(\bb{x}_{j^\prime k^\prime},\btheta_{l^*})=\frac{1}{B} \sum_{b=1}^{B}U(\bb{x}_{j^\prime k^\prime},\btheta_{l^*}^{(b)})$, $\bb{x}_{j^\prime k^\prime} \in \mathcal{A}_1$ and identify the dose  combination  $\bb{x}_{j^* k^*}=(x_{a,j^*},x_{b,k^*})$ with the highest posterior mean utility under the safety constrain $(j^*-j)+(k^*-k) \leq 1$, i.e.,
	\begin{equation}
		\bb{x}_{j^* k^*}=\argmax_{ \bb{x}_{j^\prime k^\prime}} \{ \bar{U}(\bb{x}_{j^\prime k^\prime},\btheta_{l^*}) \} \mbox{~subject to~} \bb{x}_{j^\prime k^\prime} \in \mathcal{A}_1,  \mbox{~and~}  (j^\prime-j)+(k^\prime-k) \leq 1
		\label{eq:xstar}
		\end{equation}

		[c.] If dose  combination  $\bb{x}_{j^* k^*}$ has not been used to treat any patient thus far, or all doses in $\mathcal{A}_1$ have been used to treat patients, we assign the next cohort of patients to $\bb{x}_{j^*k^*}$. However, if $\bb{x}_{j^* k^*}$ has been used and there are some untried dose combinations in $\mathcal{A}_1$, we assign the next cohort of patients to $\bb{x}_{j^* k^*}$ only if 
		\begin{equation*}
		\hat{Pr}\{U(\bb{x}_{j^*,k^*},\btheta_{l^*}) > U_0 \mid Data\} 
		\approx \frac{1}{B} \sum_{b=1}^{B} \bb{1} \{ U(\bb{x}_{j^*k^*},\btheta_{l^*}^{(b)}) > U_0 \} 
		> \left( \frac{N_2-n_2}{N_2} \right) ^ \omega
		\end{equation*}
		where $U_0$ is the lowest acceptable utility value, $n_2$ is the total number of patients that have been treated in stage II and $\omega$ is a known tuning parameter controlling how stringent the threshold is.  Otherwise, exclude $\bb{x}_{j^* k^*}$ from $\mathcal{A}_1$ and return to step b.

\noindent {\bf If $(x_{a,j},x_{b,k})$ is deemed toxic, } i.e., $Pr\{ p(\bb{x}_{jk})>p_T | Data\} > \xi$, 
	de-escalate 
	to the untried 1-degree lower doses, $\bb{x}_{j-1,k}=(x_{a,j-1},x_{b,k})$ or $\bb{x}_{j,k-1}=(x_{a,j},x_{b,k-1})$ or both. That is, if both dose combinations exist and have not been used, two cohorts of patients are recruited and assigned to $(x_{a,j-1},x_{b,k})$ and $(x_{a,j},x_{b,k-1})$, simultaneously. 
	If only one dose exists and has not been 
	used,  
	assign the next cohort of patients to this dose. If both doses exist but both have been   used,  
	terminate this cohort and do not recommend any dose for next cohort until there is a cohort newly completed. 

As seen above, we adopt a concept of 1-degree admissible neighbor $\mathcal{A}_1$ in assigning the next dose combinations when dose insertion is not needed.  \citet{Cai2014} demonstrate that this admissible neighbor and adaptive rule (step c) not only encourage the exploration of untried dose combinations to avoid the problem of trapping in suboptimal doses, but also restrict the dose escalation/de-escalation within the neighbors of the current dose, avoiding dramatic dose changes and improving the reliability of the dose-finding.

\subsection{Dose-finding Algorithm}
The AAA design is summarized in Box 1. Additional rules listed in Box 2 are for ethics and stability concern. For brevity, we use \textquotedblleft dose \textquotedblright to denote a dose combination hereinafter.  

\begin{center}
	\resizebox{\textwidth}{!}{
		\fbox{
			\parbox{7in}
			{
				\textbf{Box 1: The AAA design for phase I/II dose finding trials} \\
				\vskip -0.01 in
				The trial starts with the treatment of the first cohort of patients at the lowest dose $(x_{a,1},x_{b,1})$. Suppose that patients are being treated at dose $\bb{x}_{jk}=(x_{a,j},x_{b,k})$. A dose is deemed toxic if $Pr \{p(\bb{x}_{jk})>p_T | Data\}>\xi$, where $\xi$ is close to 1; otherwise, the dose is safe. Let $N_1$ and $N_2$ denote the maximum sample size of stage I and II, respectively. Let $n_2$ be the number of patients currently enrolled in stage II. 
				
				\begin{description}
					\item[] \textbf{Stage I Run in Period} 
					\begin{itemize}
						\item[I1] If dose $(x_{a,j},x_{b,k})$ is safe, escalate   diagonally  
						and treat the next cohort at $(x_{a,j+1},x_{b,k+1})$. If $j=k=K$, escalate 
						to $(x_{a,j+1},x_{b,K})$.
						\item[I2] Stage I is complete when either dose $(x_{a,j},x_{b,k})$ is deemed toxic or the highest dose combination $(x_{a,J},x_{b,K})$ is reached. Stage II 
						starts.
					\end{itemize}			
					\item[] \textbf{Stage II Adaptive Dose Finding}
					\begin{itemize}
						\item[II1]   Once a cohort completes follow up, collect their efficacy and toxicity outcomes.  
						\item[II2] Using the accumulated trial data on all the completers, generate MCMC posterior samples of parameters under  models  $M_1$, $M_2$, $M_3$ and $M_4$, respectively, and carry out adaptive model selection. 	  Suppose model $l^*$ is selected, denote 
						the MCMC posterior sample $\{\btheta_{l^*}^{(b)}, b=1,\ldots,B\}$ under the selected dose-efficacy model $M_{l^{*}}$, $l^{*} \in \{1,2,3,4\}$.
						\item[II3] Obtain a posterior sample $\bb{x}_{opt}$ under $M_{l^{*}}$, i.e. $\{\hat{\bb{x}}_{opt,l^*}^{(b)},  b=1,2,\cdots,B\}$, from \eqref{eq:bodc}. Then compute the posterior mean BODC, $\hat{\bb{x}}_{opt}=(\hat{x}_{a,opt},\hat{x}_{b,opt})$, from \eqref{meanbodc}, and the decision indicator $\hat{A}$ in \eqref{doseinsertion2}.
						\begin{itemize}
							\item[(a)] If $\hat{A}=1$, the new dose $\hat{\bb{x}}_{opt}$ is inserted in the trial and assigned to the next cohort. In addition, two sets of doses, $\{(x_{a,1},\hat{x}_{b,opt}), \cdots, (x_{a,J},\hat{x}_{b,opt})\}$ and $\{(\hat{x}_{a,opt},x_{b,1}), \cdots, (\hat{x}_{a,opt},x_{b,K})\}$, are inserted as well. Then go to step II1 and  
							wait for the completion of this cohort. 
							\item[(b)] If $\hat{A}=0$ and $\bb{x}_{jk}$ is safe, 
							\begin{itemize}
								\item[b1)] identify $\mathcal{A}_1$ as the set of safe neighbors of $\bb{x}_{jk}$ with degree 1.
								\item[b2)] In $\mathcal{A}_1$, identify the dose $\bb{x}_{j^*k^*}$ that has the highest posterior mean utility under the safety constraint $(j^*-j)+(k^*-k) \leq 1$, from \eqref{eq:xstar}.
								\item[b3)] If $n_{j^*k^*}=0$ or $n_{rs} \neq 0$, $\forall \ \bb{x}_{r,s} \in  \mathcal{A}_{1}$, treat the next cohort at dose $\bb{x}_{j^*k^*}$. Otherwise, \\
								if $Pr\{U(\bb{x}_{j^*,k^*},\btheta_{l^*} \mid Data) > U_0 \} > \left( \frac{N_2-n_2}{N_2} \right) ^ \omega$, treat the next cohort at  $\bb{x}_{j^*k^*}$; \\
								otherwise, remove $\bb{x}_{j^*k^*}$ from $\mathcal{A}_1$ and go to step b2.
							\end{itemize}
							\item[(c)] If $\hat{A}=0$ and $\bb{x}_{jk}$ is toxic,  de-escalate to the untried 1-degree lower doses   allowing cohort division:
							\begin{itemize}
								\item[-] If $\{j,k\geq2$ \mbox{and} $n_{j-1,k}=n_{j,k-1}=0\}$,   simultaneously enroll two cohorts of patients at both doses $\bb{x}_{j-1,k}$ and $\bb{x}_{j,k-1}$; 
								\item[-] If $\{j,k \geq 2$ \mbox{and} $n_{j-1,k}=0$ \mbox{but} $n_{j,k-1}>0\}$, or $\{ j \geq 2, k=1, \mbox{and~} n_{j-1,k}=0\}$, assign the next cohort to dose $\bb{x}_{j-1,k}$;
								\item[-] If $\{j,k \geq 2 \mbox{~and~} n_{j,k-1}=0 \mbox{~but~} n_{j-1,k}>0\}$, or $\{ k \geq 2, j=1, \mbox{and~} n_{j,k-1}=0\}$,  assign the next cohort to dose $\bb{x}_{j,k-1}$;
								\item[-] Otherwise, terminate this cohort and do not recommend any dose.
							\end{itemize}
							\item[(d)] If no dose is recommended in (a)-(c), 
							assign the next cohort to the dose $\bb{x}_{\tilde{j} \tilde{k}}$ 
							which has the highest posterior mean utility among all the existing safe doses, i.e., $\bb{x}_{\tilde{j} \tilde{k}}=\argmax_{ \bb{x}_{j^\prime k^\prime}} \{ \bar{U}(\bb{x}_{j^\prime k^\prime},\btheta_{l^*}) \}$				    
						\end{itemize}
						\item[II4] Repeat steps II1-II3 until the maximum sample size $N=N_1+N_2$ is reached.
						\item[II5] Select 
						the dose that has the highest mean utility among all tested safe doses, including 
						the newly inserted dose. 
					\end{itemize}
				\end{description}
			}
		}
	}
\end{center}

\begin{center}
	\singlespacing
	\resizebox{\textwidth}{!}{
		\fbox{
			\parbox{\hsize}{
				\textbf{Box 2: Practical rules}
				\begin{description}
					\item[--] \textbf{Rule 1 [Dose extrapolation]} The inserted new dose is not allowed to be more than twice the highest dose or less than half of the lowest dose that has been used in the trial.
					\item[--] \textbf{Rule 2  [Early Termination]} If the lowest dose $\bb{x}_{11}=(x_{a,1},x_{b,1})$ is deemed toxic, i.e., $Pr\{p(\bb{x}_{11})>p_T|Data\}>\xi$, where $\xi$ is close to 1, and no new dose is inserted, terminate the trial.
					\item[--] \textbf{Rule 3 [Dose Exclusion]} If the dose $\bb{x}_{jk}=(x_{a,j},x_{b,k})$ is deemed toxic, i.e., $Pr\{p(\bb{x}_{jk})>p_T|Data\}>\xi$, where $\xi$ is close to 1, exclude doses $\{(x_{a,j^\prime},x_{b,k^\prime}):j^\prime=j,j+1,\ldots,J,\ k^\prime=k,k+1,\ldots,K\}$, i.e. these doses will never be used in the trial again.
					\item[--] \textbf{Rule 4 [No Skipping Dose]} Restrict the escalation to 1 level increment,  
					i.e. there is no skipping in the escalation. Particularly, if the new dose intended for insertion is higher than any unexplored dose, pause the insertion and go to step II3(b)-II3(d).
				\end{description}
			}
		}
	}
\end{center}

\section{Simulation}  \label{results}
\subsection{Simulation setup}{\label{simu-setup}}
We consider the motivating trials combining two agents, a MEK inhibitor and a PIK3CA inhibitor, 
each with $4$ dose levels. The maximum sample size is 96 and cohort size is 3. We investigate 10 different scenarios, and all scenarios assume a true 
linear or 
quadratic logistic model for both agents in the dose-efficacy relationship, as shown in Figure \ref{fig:scenario}. For each scenario, 1,000 simulated trials are conducted. In the proposed design, we set the MTD toxicity threshold $p_T=0.3$, the credible level threshold $C\%=0.90$, the lowest acceptable utility value $U_0=0.1$ and the tuning parameter $\omega=2$. The probability threshold $\xi=0.95$ for the practical rule and safe requirement. Regarding the utility function, we assume that the two toxicity and efficacy rate pairs, $(0,0.45)$ and $(0.3,0.85)$ have the same utility value $0.3$. As a result, we obtain the estimated $\bb{\hat{\eta}}=(0.396,0.385,1.280,-0.385)$.

For MCMC computation, we adopt a 
standard  
random walk Metropolis-Hasting algorithm. 
And for each chain, 10,000 MCMC samples are drawn with a burn-in size of 5,000 iterations. 
The MCMC mixed fast and well with no sign of convergence problems.

For comparison, we apply 
the design in \citet{Cai2014}. For fairness, we slightly modified this design by using the utility function rather than efficacy probability for defining the dose in admissible dose set $\mathcal{A}$. This typically improved the performance of their design based on our experience. 
There is no dose insertion and model selection in their algorithm. Therefore, we only compare the dose allocation and dose selection. 
To demonstrate the benefit of adaptive cohort division, we turn off the ACD procedure in AAA and apply 
a single cohort algorithm, in which we randomly select one dose in step IIc if two doses are available for ACD. 
See Web Appendix B for details of the simulation scheme for patients enrollment and follow-up. 

\subsection{Operating characteristics} \label{oprating}
The operating characteristics of the proposed algorithm for Scenarios 1-6 are summarized in Table \ref{tab:sc1-6} with seven sections per scenario. 
Section 1 gives a brief description of the true dose response and the need for insertion. Sections 2 and 3 provide the  mean  (standard deviation) of the posterior means across the 1,000 simulated trials for the regression parameters in the dose-toxicity and dose-efficacy models, respectively. Section 4 summarizes the model selection 
frequencies. 
Section 5 lists the prespecified dose levels, and the true toxicity and efficacy probabilities of all the doses. Section 6 presents the detailed operating characteristics of the AAA  design  in terms of patient allocation and dose selection. 
Lastly, Section 7 presents the results of the modified algorithm from \citet{Cai2014}. For brevity, we present simulation results for Scenarios 7-10 in Web Table 2.

\begin{figure}
	\begin{center}
		\centerline{\includegraphics[width=1\textwidth]{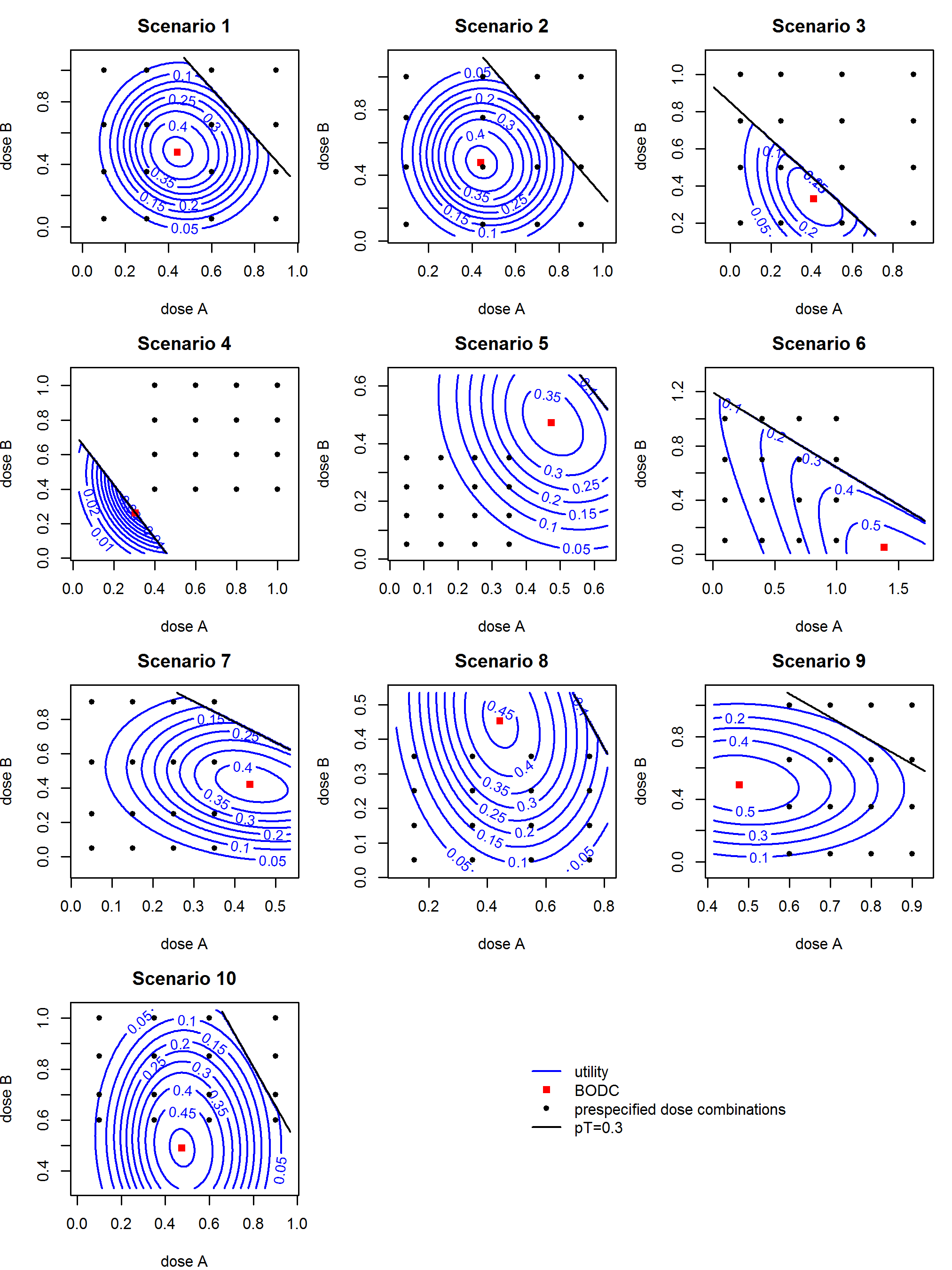}}
		\caption{ \small Ten scenarios in the simulation study. Black dots represent the prespecified dose combinations. Red dots marks the true BODC. 
			Blue contour: dose-utility contour. Black line:   dose combinations with toxicity probability $p_T$.
			Any dose combinations beyond the $p_T$ line with higher toxicity probability are of 0 utility.} 
		\label{fig:scenario}
	\end{center}
\end{figure}

In Scenario 1, the efficacy rates firstly increase and decrease later with both agents. The true BODC $\bb{x}_{opt}=(0.441,0.476)$ 
is bracketed by 
doses $(0.3,0.35)$, $(0.3,0.65)$, $(0.6,0.35)$ and $(0.6,0.65)$. 
Therefore, new dose should be inserted (Figure \ref{fig:scenario}). From Table \ref{tab:sc1-6}, we see that the AAA design inserts new doses with mean $(0.430,0.448)$ in 59.3\% of the simulated trials. 
The utility of the mean inserted dose combination under the true model is $0.465$, which is higher than that of all prespecified dose combinations. Among all the patients, 25.9\% are treated at the inserted new dose combinations. At the end of 55.9\% of the simulated trials, an inserted dose is claimed to be the BODC. The mean selected BODC at the end of the trial is $(0.456,0.464)$, close to the true BODC. Also, 90.5\% of the trials correctly choose the quadratic logistic regression for both agents for dose-efficacy curve at the end, and the posterior sample means of $\bbeta$ are close to the true values. 
Section 7 shows the selection percentages and patient allocation based on \citet{Cai2014}. The results are reasonable as most patients are allocated to the four doses that surround the true BODC. However, since their design dose not allow dose insertion, it cannot correctly identify the true BODC.

Scenario 2 has a similar pattern of dose-toxicity and dose-efficacy curves with Scenario 1. From Figure \ref{fig:scenario}, 
there is no need for insertion. The AAA  design  only inserts new doses in 15.5\% of trials. 
The mean selected dose combination is $(0.454,0.466)$, close to the true BODC $(0.442,0.477)$. 

Scenario 3 reflects a setting where 
only a few doses are tolerable while others are overly toxic (Figure \ref{fig:scenario}). 
Because the utility of the existing doses $\bb{x}_{31}=(0.55,0.2)$ and $\bb{x}_{22}=(0.25,0.5)$ are 0.240 and 0.217, respectively, which are not much different from the utility of true BODC $(0.41,0.332)$, 0.297, only 29.6\% of the trials insert new doses. 
Note that in this scenario, AAA selects the true quadratic logistic model for the dose-efficacy 73\% of the times, a smaller percentage than Scenarios 1 and 2. 
This is because only a few patients (22.7\% in total) are assigned to the doses at the upper right corner of the dose matrix (shown in Figure \ref{fig:scenario}) due to their high toxicity. 
Therefore, there is insufficient amount of information for better model selection in this scenario.

Scenario 4 is a situation when all combinations are higher than MTD, and hence 8.6\% of the trials are terminated at an early stage according to Practical Rule 2. Among completed trials, the proportion of selecting the true quadratic model for efficacy is 67.8\%. New dose combinations with a mean $(0.354,0.354)$ are inserted among 92.5\% of the completed trials, and 96.7\% end with selecting new dose combinations as BODC. The mean selected dose combination is $(0.242,0.245)$, which is lower than the MTD and close to the true BODC $(0.304,0.262)$. This scenario demonstrates the important safety feature of the proposed design. Also, AAA not only stops the trial early due to toxicity, it performs desirable dose insertion below all the prespecified doses and identifies the correct BODC.

Unlike the previous four scenarios, Scenario 5 presents a situation where the prespecified dose matrix only covers the bottom-left corner of the quadratic dose-efficacy curve. Thus, the efficacy grows with both agents and the true BODC $(0.474,0.472)$ locates at the upper-right corner beyond the dose matrix. The insertion rate is 11.1\% and the mean selected BODC is $(0.358,0.345)$, which is poorly estimated. And all four dose-efficacy models are selected at similar rates. 
There are two reasons. First, because the prespecified doses do not cover a wide range of dose response surface, data on these doses could not provide a good estimate of the entire curve. Second, with a relatively small sample size $N=96$, the simulated trials often run out of patients before the BODC combination is reached. If a large $N$ is allowed, the dose matrix could be extrapolated well and the dose insertion algorithm would perform better (results not shown). 

Scenario 6 assumes that the true dose-efficacy is a linear logistic model. About 32.2\% of the simulated trials insert new doses, which is as expected, and the linear model is selected at a relatively high rate, 70.3\%. A total of 28.7\% of trials select the new doses as the BODC with the mean $(1.002,0.341)$.

In Scenarios 7-10 (see Web Table 2), the dose-efficacy curves increase first and fall later with one agent, but are monotone with the other agent. In addition, the true BODC are all located outside the prespecified dose matrix， at the right-side, top-side, left-side and bottom-side regions, respectively. In all four scenarios, insertion is needed, and the average insertion rates are 44.2\%, 29.7\%, 88.2\% and 79.9\%, respectively. The mean selected doses are close to the true BODCs in all four scenarios.

Comparing to the results of \citet{Cai2014} in Section 7 of all scenarios, we can see that the doses close to the true BODC are assigned more patients and selected more frequently.   This shows that their design performs well in general. However, when encountering situations where dose insertion is needed, their design is unable to select the true BODC due to the lack of dose insertion.

\subsection{Time duration}
Table \ref{tab:time} demonstrates 
the benefit of ACD in shortening trial duration. In particular, step IIc in  Box 1  is expected to 
speed up the trial process and reduce the time duration. It can be seen from Table \ref{tab:time} that about 100 days or more can be saved across most scenarios with the ACD procedure. The trial is never longer with ACD than without ACD. 
Scenarios with more toxic doses result in more reduction of trial time. 
For example, for Scenario 3, the trial duration is reduced by about 400 days. However, if all doses are overly toxic, the reduction of time duration is negligible, since cohort division is not allowed for inserted doses.  
The performance of multiple cohorts and single cohorts are almost the same in terms of the mean selected BODC, patients allocation, and the percentage of being selected as the BODC. 

\subsection{Sensitivity to sample size}{\label{samplesize}}
Lastly, to evaluate the effect of sample size, we apply the algorithm with a smaller sample size of 66 on Scenarios 1, 4 and 5. Results are summarized in Web Table 3. We find the that the reduction of sample size has larger influence on Scenario 1 than Scenarios 4 and 5. Specifically, for Scenario 1, the insertion rate is reduced to 42\% 
(from 55.9\%) 
and the proportion of selecting the correct quadratic model is declined to 87.5\% 
(from 90.5\%), 
although the mean inserted dose combination $(0.44,0.455)$, and the mean selected BODC $(0.460,0.466)$, are still close to the truth.   Overall, the AAA perform reasonably well with a significant reduction of sample size. 

\section{Discussion}  \label{discussion}
We propose a new Bayesian adaptive dose insertion design for dual-agents phase I/II oncology trials. The dose-insertion procedure based on both efficacy and toxicity enables us to locate more desirable dose combinations. Bayesian model selection during the trial allows the dose-efficacy relationship to be adapted between linear and quadratic logistic models. 
The model selection has been shown be important \citep{guo2015teams} in dose insertion and maintaining a high efficiency of the dose-finding trial. 
The adaptive cohort division speeds up the trial process and shortens the time duration in most scenarios. Simulation results show that the proposed design has superior operating characteristics. 

The AAA design is a utility-based method. Clearly, the performance heavily depends on the definition of utility. Although we choose utility as a multiplication of a linear truncated function (safety utility) and an exponential function (efficacy) in this paper, one can use other reasonable alternative utilities. 
Choosing an appropriate utility function has to be done for individual trials and through discussion between clinicians and statisticians. For different diseases and drugs, different trade-off between efficacy and toxicity might be allowed. Nevertheless, changes on the utility function is a separate topic and does not affect the overall statistical design illustrated in Boxes 1 and 2.  In other words, the algorithm presented therein is expected to find the optimal dose combination with high likelihood based on the defined utility, however it is defined.  

In our models, we do not include an interaction effect $\beta_5 x_a x_b$ for the two agents for a couple of reasons. First, our current clinical trial cannot afford to recruit a large number patients, say a few hundreds, which is typically needed for estimating interaction terms in a regression model reliably. Second, it has been demonstrated that for the purpose of dose finding, a local fit with a working model of the response surface does not affect much the efficiency of dose finding. That is, even when models are mis-specified (by not including a true interaction term), the dose-finding decisions are not affected severely and still lead to reasonable performance. This has been shown in \citet{Cai2014} and \citet{wang2005two} in the context of drug combination trials. To see this, we simulated two scenarios consisting of the interaction term $\beta_5 x_a x_b$ (Web Figure 2) and summarize results in Web Table 4. We can see that the AAA design performs well in finding the optimal dose combination and allocating patients to desirable dose combinations. The inference on the the parameter of the interactive term $\beta_5$ is biased.    Typically, accurate estimation of $\beta_5$ requires a large sample size and well-placed dose combinations across the dose-response surface. In our current example, we only have $4\times 4$ dose levels for two agents, which leads to the difficulty in capturing the rotation of the efficacy curve, i.e., $\beta_5$.  We found that when we increased to a combination of $6\times 6$ and $7\times 7$ dose levels, the interaction can be well estimated (results are not shown).

\section*{Supplementary Materials}
Appendices, Tables, and Figures, referenced in Sections \ref{AMS}, \ref{likelihood}, \ref{simu-setup}, \ref{oprating} and \ref{discussion}, are available with this paper at the online website. 
\vspace*{-8pt}

\section*{Acknowledgements}
{ Yuan Ji's research is partly supported by NIH R01 132897-06.} \vspace*{-8pt}

\bibliographystyle{rss} 
\bibliography{bib_2dteams}

\clearpage
\newpage

\clearpage
\newpage
\begin{center}
	\LTcapwidth=\textwidth
	\begin{longtable}{ccccc}
		\caption{Median probability model selection rules.} \\ 
		\endfirsthead
		\multicolumn{5}{c}{\tablename\ \thetable\ (continued)}
		\endhead
		\hline \hline 
		\emph{$p_3$} & \emph{$p_4$} & \emph{$\beta_3$} & \emph{$\beta_4$} & \emph{The Selected Dose-Efficacy Model}\\
		\hline
		$<1/2$ & $<1/2$ & $=0$ & $=0$ & linear logistic model, $M_1$ \\ 
		$\geq 1/2$ & $<1/2$ & $\neq 0$ & $=0$ & Quadratic for agent A, $M_2$ \\ 
		$< 1/2$ & $\geq 1/2$ & $=0$ & $\neq 0$ & Quadratic for agent B, $M_3$ \\ 
		$\geq 1/2$ & $\geq 1/2$ & $\neq 0$ & $\neq 0$ & Quadratic for both agents A and B, $M_4$ \\ 
		\hline \hline 
		\label{tab:mpm}
	\end{longtable}
\end{center}

\clearpage
\newpage
\begin{center}
	\fontsize{8}{10}\selectfont\rm
	\LTcapwidth=\textwidth
	\setlength\LTleft{-0.5in}
	\setlength\LTright{-0.5in plus 1 fill}
	\begin{longtable}{ccccccc}		
		\caption{Simulation results for scenarios 1-6. For each scenario, 1,000 trials are conducted on computer and the operating characteristics are summarized in seven sections. Section 1 gives a brief description of the true dose response and the need for insertion or not. Section 2 and 3 provide the average (standard deviation) of the posterior means across 1,000 simulated trials for the regression parameters. Section 4 summarized the model selection. Section 5 presents the true toxicity and efficacy probability at each dose combination and Section 6 presents the detailed operating characteristics in terms of patient allocation and dose selection. Section 7 demonstrates the results of the design in \citet{Cai2014}. }\\ 
		\endfirsthead
		\multicolumn{7}{c}{\tablename\ \thetable\ (continued)}
		\endhead
		\csname @@input\endcsname sc1_own.tex 
		\csname @@input\endcsname sc2_own.tex 
		\csname @@input\endcsname sc3_own.tex 
		\csname @@input\endcsname sc4_own.tex 
		\csname @@input\endcsname sc5_own.tex 
		\csname @@input\endcsname sc11_own.tex 
		\label{tab:sc1-6}
	\end{longtable}
\end{center}

\clearpage
\newpage
\begin{center}	
	\LTcapwidth=\textwidth	
	\begin{longtable}{ccccc}
		\caption{Time duration comparison between the   adaptive cohort division 
			algorithm and the single cohort under the AAA. Entries are simulated trial duration is days for two-agent trials of sample size 96 patients. }\\ 
		\endfirsthead
		\multicolumn{5}{c}{\tablename\ \thetable\ (continued)}
		\endhead
		
		\csname @@input\endcsname time.tex
		\label{tab:time}
		
	\end{longtable}
\end{center}

\end{document}